\documentclass[12pt]{article}

\usepackage[margin=1in]{geometry}
\usepackage{setspace}
\usepackage{amsmath,amssymb,amsthm}
\usepackage{graphicx}
\usepackage{booktabs}
\usepackage{natbib}
\usepackage{hyperref}
\usepackage{xcolor}
\usepackage{tikz}
\usepackage{float}
\usetikzlibrary{positioning,arrows.meta,shapes,calc}

\usepackage{soul}
\usepackage[normalem]{ulem} 

\definecolor{hlgreen}{RGB}{144,238,144}
\definecolor{hlpink}{RGB}{255,182,193}
\definecolor{hlblue}{RGB}{173,216,230}
\definecolor{hlorange}{RGB}{255,200,100}

\sethlcolor{yellow}

\newtheorem{definition}{Definition}
\newtheorem{assumption}{Assumption}

\hypersetup{
    colorlinks=true,
    linkcolor=black,
    citecolor=black,
    urlcolor=blue
}

\title{\textbf{Beyond Interaction Effects: Two Logics for Studying Population Inequalities}}

\author{
    {\Large Adel Daoud}\thanks{I thank my colleagues at the AI and Development Lab for valuable discussions. \textbf{Use of Generative AI:} The author used generative AI tools (Claude, Anthropic; GPT 5.2 Pro; Gemini Pro 3) to assist with literature review, citation verification, and manuscript preparation. All AI-generated content was reviewed, verified, and edited by the author, who takes full responsibility for the final manuscript.}\\
    {Institute for Analytical Sociology, Linköping University, Sweden}\\
    {Division of Data Science and AI, Chalmers University of Technology, Sweden}\\
    {The AI and Development Lab, \url{www.aidevlab.org}} \\
    {adel.daoud@liu.se}
}

\date{\today \\ \small Draft v2.0}

\begin{document}

\maketitle

\newpage
\begin{abstract}
\noindent
When sociologists and other social scientist ask whether the return to college differs by race and gender, they face a choice between two fundamentally different modes of inquiry. Traditional interaction models follow deductive logic: the researcher specifies which variables moderate effects and tests these hypotheses. Machine learning methods follow inductive logic: algorithms search across vast combinatorial spaces to discover patterns of heterogeneity. This article develops a framework for navigating between these approaches. We show that the choice between deduction and induction reflects a tradeoff between interpretability and flexibility, and we demonstrate through simulation when each approach excels. Our framework is particularly relevant for inequality research, where understanding how treatment effects vary across intersecting social subpopulation is substantively central.

\vspace{1em}
\end{abstract}

\newpage
\setcounter{page}{1}

\section{Introduction}

Consider a policy question that lies at the heart of stratification research: Who benefits most from attending college \citep{brand2010, xie2013}? The average return to a bachelor's degree is well documented---approximately 50-80 percent higher lifetime earnings compared to high school graduates. But this average conceals enormous variation. Some individuals experience transformative gains while others see modest returns or even negative effects once opportunity costs are considered. A Black woman from a low-income family, a White man whose parents both hold graduate degrees, and a first-generation immigrant student occupy different positions in social space, and we have good reason to expect that college affects their life trajectories differently.

This variation in causal effects---what applied researchers tend to call \textit{interaction effects} or \textit{effect heterogeneity} and methodologists call \textit{conditional average treatment effect}---is not merely a statistical nuisance to be controlled. It is often the phenomenon of primary substantive interest. Understanding how the effect of education varies by race and class speaks directly to theories of cumulative advantage and stratification. Knowing which students benefit most from a particular intervention determines whether that intervention reduces or exacerbates existing inequalities. Effect heterogeneity research transforms causal inference from a search for universal laws into an investigation of contingent mechanisms operating differently across social contexts.

Yet researchers face a methodological divide when seeking to estimate this heterogeneity. The traditional approach relies on multiplicative interaction terms in regression models, formalized decades ago and still dominant in applied research \citep{aikenwest1991, brambor2006}. This approach requires researchers to specify in advance which variables moderate the treatment effect and how they combine. A study testing whether the return to college differs by race would include a college-by-race interaction term; a study examining whether this racial difference itself varies by gender would add a three-way interaction. The logic is fundamentally deductive: theory generates hypotheses about effect modification, and data test these predictions.

An alternative has emerged from recent advances in machine learning \citep{brand2023, molinagarip2019, daoud2021melting}. Methods like causal forests \citep{atheywager2018}, meta-learners \citep{kunzel2019}, and Bayesian additive regression trees \citep{hill2011} can discover complex patterns of heterogeneity directly from data without requiring researchers to specify functional forms in advance. Rather than testing whether race moderates the college effect, these algorithms search across all observed characteristics---race, gender, family income, parental education, test scores, region, and their myriad combinations---to find where treatment effects concentrate. The logic is fundamentally inductive: patterns emerge from data, and researchers interpret what algorithms find, and develops theory from these interpretations. 

This article develops a framework for navigating between these two modes of discovery \citep{daoud2023hdsr}. Our central argument is that the choice between deductive and inductive approaches reflects a fundamental tradeoff between theory testing and theory development, as well as interpretability and flexibility, and that neither mode dominates the other across all research contexts. We demonstrate through simulation when traditional interaction models outperform machine learning methods and when the reverse holds. We connect this methodological discussion to substantive questions in inequality research, where understanding how effects vary across intersecting categories of race, class, and gender is central to the sociological enterprise \citep{daoud2016satpov, daoud2015poverty}.

The article proceeds as follows. Section 2 develops the distinction between deductive and inductive approaches to effect heterogeneity, showing how this choice structures the research process from start to finish. Section 3 clarifies the conceptual foundations---what exactly we mean by heterogeneous effects and what different estimands capture. Section 4 reviews methods for estimating effect heterogeneity, treating traditional and machine learning approaches within a unified framework. Section 5 presents simulation results comparing these methods across scenarios designed to illuminate when each excels. Section 6 discusses implications for inequality research. We conclude with recommendations for practice.

\section{Two Modes of Discovery: Deductive versus Inductive}

The distinction between deductive and inductive approaches to effect heterogeneity is not merely technical. It reflects fundamentally different philosophies about how scientific knowledge accumulates and what role theory plays in empirical research \citep{morganwinship2015}. Understanding this distinction is essential for making principled methodological choices.

Deductive approaches begin with theory. The researcher draws on existing knowledge to formulate specific hypotheses about which variables should moderate treatment effects and in what direction \citep{aikenwest1991}. A sociologist studying returns to education might theorize, based on cumulative advantage frameworks, that college provides greater benefits to students from disadvantaged backgrounds because it offers resources and credentials they could not otherwise access \citep{brand2010}. Alternatively, drawing on theories of credential inflation, the researcher might hypothesize that returns are higher for students from advantaged backgrounds because they can better convert educational credentials into labor market outcomes through network connections and cultural capital. Either way, the theoretical argument specifies the interaction to be tested: the researcher estimates a model including a college-by-family-background interaction term and examines whether the coefficient is statistically significant and in the predicted direction \citep{brambor2006}.

This deductive approach has considerable strengths. It maintains a tight connection between theory and evidence, ensuring that empirical findings speak directly to theoretical questions \citep{morganwinship2015}. It controls multiple testing problems by pre-specifying hypotheses rather than searching through data for significant patterns. It produces interpretable results: a statement like ``the return to college is 0.15 log points lower for students whose parents hold college degrees'' is immediately meaningful to sociological audiences. And it forces researchers to think carefully about mechanisms before examining data, which can prevent post-hoc rationalization of atheoretical findings \citep{gelman2018}.

But deductive approaches also have significant limitations. They can only detect heterogeneity patterns that researchers think to specify \citep{athey2019}. A study testing race-by-college and gender-by-college interactions will miss a four-way race-by-gender-by-class-by-region interaction, even if this pattern is strong in the data. This matters particularly for intersectionality research, where the whole point is that social positions are defined by combinations of attributes that may interact in ways not predictable from their separate effects \citep{mccall2005, bauer2021}. Deductive approaches also impose functional form assumptions---typically linearity---that may not match true data-generating processes. When the effect of college varies with parental income in a nonlinear way, a linear interaction term will at best approximate the true pattern and at worst miss it entirely \citep{hainmueller2019}.

Inductive approaches reverse the logic. Rather than beginning with theory, the researcher begins with data and asks algorithms to discover patterns of heterogeneity \citep{atheywager2018}. Causal machine learning methods like causal forests partition the covariate space to maximize treatment effect heterogeneity across partitions, without requiring the researcher to specify which variables matter or how they combine \citep{chernozhukov2024causalml}. The algorithm might discover that treatment effects are concentrated among individuals with a particular combination of characteristics---say, Black women from low-income families with high test scores living in rural areas---that no researcher would have thought to specify in advance.

This inductive approach also has considerable strengths. It can discover unexpected patterns that deductive approaches would miss \citep{brand2021}. It handles high-dimensional covariate spaces where the number of possible interactions exceeds what any researcher could enumerate. It makes no assumptions about functional form, allowing effects to vary in arbitrarily complex ways with covariates \citep{hill2011}. And it is particularly valuable in exploratory research contexts where existing theory provides limited guidance about where heterogeneity might reside \citep{molinagarip2019}.

But inductive approaches carry their own limitations. They risk overfitting---finding patterns that reflect noise in the sample rather than true population heterogeneity \citep{athey2019}. They may produce results that can be difficult to interpret and communicate: a causal forest generates many sub-group effect estimates, but explaining why one group's estimated effect is 0.3 and another's is 0.7 requires additional theorizing. They are susceptible to multiple testing problems because algorithms implicitly examine many possible interaction patterns \citep{gelman2018}. And they can generate findings that lack theoretical grounding, making it unclear what mechanisms produce the observed heterogeneity.

The choice between deductive and inductive approaches is not simply a matter of method but of research philosophy \citep{daoud2023hdsr}. Deduction emphasizes confirmation: testing whether theoretically predicted patterns appear in data. Induction emphasizes discovery: finding patterns that might inform new theoretical developments. Both modes are essential to scientific progress \citep{daoud2021melting}. Purely deductive research risks becoming an exercise in confirming researchers' prior beliefs while missing unexpected phenomena. Purely inductive research risks generating atheoretical pattern descriptions that do not cumulate into systematic knowledge.

A promising strategy combines both modes in sequence. The researcher begins inductively, using machine learning to discover which variables drive heterogeneity and how they combine. Then, guided by these discoveries, the researcher formulates theoretical interpretations and tests them deductively in new samples or with more targeted analyses. This two-stage approach preserves the strengths of each mode while mitigating their limitations. We return to this strategy after reviewing the methods in detail.

\section{What Is Effect Heterogeneity?}

Before comparing estimation approaches, we must clarify what we seek to estimate \citep{lundberg2021what}. Effect heterogeneity refers to systematic variation in causal effects across units or subgroups. This section develops the conceptual foundations, leaving formal definitions to Appendix A.

The most common target in causal inference is the average treatment effect, which answers the question: what is the effect of treatment (exposure, action, or policy), averaging over the population? This quantity is useful for determining whether an intervention ``works'' in some aggregate sense, over a population. If the average effect of a job training program on earnings is positive and statistically significant, we have evidence that the program benefits participants on average. But this average conceals potentially important variation. Some participants might experience large gains while others experience no benefit or even harm. The average effect tells us nothing about this distribution.

The conditional average treatment effect extends this by asking: what is the treatment effect for individuals with particular characteristics? This quantity answers questions like ``what is the effect of college for Black women from low-income backgrounds?'' The conditioning variables define subgroups, and the conditional average treatment effect gives the average effect within each subgroup. This is the primary target for most heterogeneity research. It connects directly to questions about disparities: if the conditional average treatment effect of college is larger for White students than for Black students, educational expansion alone will not close racial income gaps \citep{lundberg2025gapclosing}.

The most granular quantity is the individual treatment effect (ITE), which asks: what is the treatment effect for this specific person? This quantity is fundamentally unobservable because we never see both potential outcomes for the same individual---the ``fundamental problem of causal inference'' \citep{holland1986}. We observe what happened to a person who attended college, but we cannot observe what would have happened to that same person had they not attended. The ITE for person $i$ is $\tau_i = Y_i(1) - Y_i(0)$, but since we observe only one of these potential outcomes, $\tau_i$ remains forever unknown. However, what we can do is to estimate it \cite{daoud2021melting}.

This distinction between CATE and ITE deserves careful attention because the two quantities answer different questions and require different assumptions for identification. CATE---the conditional average treatment effect---averages over individuals with identical observed covariates, $X = x$. When we estimate $\tau(x) = \mathbb{E}[Y(1) - Y(0) \mid X = x]$, we are asking about the average effect for everyone who shares a particular covariate profile. This quantity can be identified under standard causal assumptions (conditional ignorability, overlap, SUTVA) without knowing anything about individuals' unobserved characteristics. In contrast, ITE asks about a specific individual's treatment effect, which depends on that individual's unique combination of observed and unobserved characteristics \citep{vegetabile2021}.

Pearl's structural approach to counterfactual inference \citep{pearl2009} provides a formal framework for thinking about ITE. In his framework, computing an individual counterfactual requires three steps. First, \textit{abduction}: given what we observed about this individual, infer the values of unobserved factors (exogenous variables or noise variable, $U$) that must have produced their observed outcome. Second, \textit{action}: perform a surgical intervention on the structural model, replacing the natural treatment mechanism with a hypothetical assignment. Third, \textit{prediction}: propagate this intervention through the modified causal model to compute what the outcome would have been under the counterfactual treatment. This three-step process reveals why ITE is fundamentally more demanding than CATE: it requires knowledge of unobserved variables that, by definition, we do not measure \citep{pearl2009, hernan2020}. Knowledge that we can acquire only when our structural assumptions are fully specified (knowing the DAG, all the functional relationships, and inverting it). CATE, on the other hand, only requires that our DAG contains the correct variables regarding the outcome and treatment, but not all internal relationships among all other variables in the DAG.   

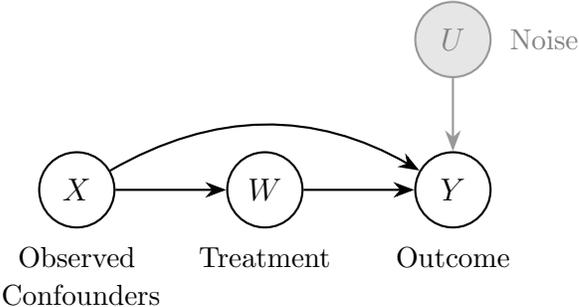
\begin{figure}[htbp]
\centering
\begin{tikzpicture}[
    node distance=2.5cm,
    observed/.style={circle, draw=black, thick, minimum size=1cm, fill=white},
    unobserved/.style={circle, draw=black!40, thick, minimum size=1cm, fill=gray!20},
    arrow/.style={-{Stealth[scale=1.2]}, thick}
]
\node[observed] (X) {$X$};
\node[observed] (W) [right of=X] {$W$};
\node[observed] (Y) [right of=W] {$Y$};
\node[unobserved] (U) [above of=Y, yshift=-0.5cm] {\textcolor{black!50}{$U$}};

\draw[arrow] (X) -- (W);
\draw[arrow] (X) to[bend left=30] (Y);
\draw[arrow] (W) -- (Y);
\draw[arrow, black!40] (U) -- (Y);

\node[below=0.1cm of X, text width=2cm, align=center] {\small Observed\\Confounders};
\node[below=0.1cm of W, text width=2cm, align=center] {\small Treatment};
\node[below=0.1cm of Y, text width=2cm, align=center] {\small Outcome};
\node[right=0.1cm of U, text width=2.5cm, align=left] {\small \textcolor{black!50}{Noise}};
\end{tikzpicture}
\caption{Structural causal model distinguishing CATE from ITE. Observed confounders $X$ affect both treatment $W$ and outcome $Y$. Unobserved factors $U$ (shown greyed) also influence outcomes. CATE can be identified by adjusting for $X$; ITE would require knowledge of individual-specific values of $U$, which we do not observe.}
\label{fig:dag}
\end{figure}

Figure~\ref{fig:dag} illustrates this distinction using a directed acyclic graph (DAG). The diagram shows a structural causal model where treatment $W$ affects outcome $Y$, confounders $X$ affect both treatment and outcome, and unobserved factors $U$ also influence the outcome. To estimate CATE, we need only adjust for the observed confounders $X$---standard causal inference methods accomplish this. But to estimate ITE, we would need to know the value of $U$ for each individual, which requires inverting the outcome equation (abduction) using information we do not have. The unobserved $U$ represents everything about an individual that affects their outcome beyond what we measure: their unrecorded motivation, aptitude, family circumstances, luck. CATE averages over this unobserved heterogeneity; ITE attempts to pin it down.

In practice, our best approximation to the individual effect is the conditional average effect given all observed characteristics, which assumes that individuals with identical observed covariates have identical treatment effects---an assumption that is almost certainly false but often necessary. When machine learning methods like causal forests produce ``individual-level'' effect estimates, they are technically producing CATE estimates at the finest granularity permitted by the data, not true ITEs \citep{kunzel2019}. This distinction matters for policy: targeting interventions based on estimated individual effects requires acknowledging the uncertainty that comes from unobserved heterogeneity.

A related conceptual distinction---between \textit{effect heterogeneity} and \textit{causal interaction}---further clarifies what different analyses identify \citep{vanderweele2009interaction, vanderweele2009dag}. Effect heterogeneity (also called effect modification) refers to a situation where a single treatment's effect varies with the level of some covariate. When we find that the effect of college on earnings differs by race, we have identified effect heterogeneity: the treatment is college, and race modifies its effect. Importantly, this modification need not be causal. Race itself may not cause the differential return to college; rather, race may be associated with unmeasured factors (discrimination, network access, occupational sorting) that produce the differential. Identifying effect heterogeneity tells us \textit{for whom} effects are larger or smaller but does not necessarily tell us \textit{why}.

Causal interaction, by contrast, involves two (or more) treatments whose combined effect differs from the sum of their separate effects \citep{vanderweele2009interaction}. If we could experimentally assign both college attendance and job training, causal interaction would exist if the effect of receiving both treatments exceeds (or falls short of) the sum of their individual effects. This is a stronger claim requiring that both variables be manipulable treatments with defensible causal interpretations. Much of what applied researchers call ``interaction effects'' is technically effect modification rather than causal interaction---a distinction with important implications for what conclusions can be drawn \citep{nilsson2021}. Throughout this article, we focus primarily on effect heterogeneity, acknowledging that the covariate moderating treatment effects may not itself have a causal interpretation.

These estimands form a hierarchy from coarse to fine. The average treatment effect is a single number summarizing the entire population. The conditional average treatment effect is a function of covariates, potentially taking different values for each combination of characteristics. The individual treatment effect is a value for each unit, the finest granularity possible. Moving down this hierarchy increases granularity but also increases estimation difficulty and decreases interpretability. A key insight of our framework is that the choice of estimand should precede the choice of method: different research questions call for different levels of granularity.

For causal identification of any of these quantities, standard assumptions must hold \citep{morganwinship2015}. Treatment assignment must be independent of potential outcomes conditional on observed covariates---meaning we must measure all confounders. Every combination of covariate values must have some treated and some untreated individuals---ensuring we can make comparisons within each subgroup. And one person's treatment must not affect another person's outcome---ruling out interference effects. Machine learning methods do not relax these assumptions. A causal forest with rich covariates still produces biased estimates if unmeasured confounding exists. The appeal of machine learning lies in flexible estimation of complex relationships, not in weaker identification requirements.

\section{Methods in Practice}

With estimands clarified, we turn to estimation methods. Rather than treating traditional and machine learning approaches as separate worlds, we present them within a unified framework highlighting their shared goals and contrasting assumptions.

The traditional approach to effect heterogeneity uses multiplicative interaction terms in regression models, often two but seldom more than three, as interpretation complexity starts mounting. The researcher specifies a model in which the outcome depends on treatment, covariates, and the product of treatment and selected covariates. The coefficient on this product term captures how the treatment effect changes with the interacting variable. For example, a model including college, race, and their product yields separate effect estimates for each racial group, with the interaction coefficient measuring the difference between groups. This approach extends naturally to multiple interactions: adding gender and a three-way college-by-race-by-gender term yields effect estimates for each race-gender combination.

The interaction approach with linear models embodies deductive logic perfectly. Researchers must specify which interactions to include based on theoretical expectations. The model tests whether these specified patterns exist in data. In linear models, results are immediately interpretable: a coefficient of 0.15 on the college-by-race interaction means the return to college is 0.15 log points higher for the omitted racial category. In generalized linear models such as logistic regression, interpretation becomes more complex because interaction coefficients on the log-odds scale do not translate directly to interactions on the probability scale, and comparisons across models with different covariates can be misleading due to unobserved heterogeneity \citep{mood2010logistic}. But regardless of whether one uses OLS or generalized linear models, the approach imposes strong assumptions. Effects must vary linearly (or linearly in the transformed scale) with continuous moderators unless the researcher specifies otherwise. Unspecified interactions are assumed zero. Sample size requirements grow rapidly with interaction order, with reliable detection of three-way interactions typically requiring thousands of observations \citep{gelman2018, hainmueller2019}.

Machine learning methods approach the same problem differently. Rather than specifying interactions, algorithms discover them \citep{brand2021, daoud2021melting}. Causal forests \citep{atheywager2018, athey2019}, the most widely used method, adapt the random forest algorithm for causal inference. The algorithm builds many decision trees, each partitioning the covariate space to maximize treatment effect heterogeneity across partitions. Averaging across trees produces smooth estimates of the conditional average treatment effect for each individual. The algorithm incorporates several innovations that distinguish it from standard prediction forests: it uses separate subsamples for tree construction and effect estimation to reduce overfitting, it residualizes outcomes to improve efficiency, and it produces valid confidence intervals despite the complexity of the underlying model.

Meta-learners offer an alternative machine learning approach \citep{nie2021, kennedy2023}. These strategies transform causal effect estimation into standard prediction problems, allowing any supervised learning algorithm to serve as the underlying engine. The simplest version fits a single model predicting outcomes from treatment and covariates, then computes effect estimates as the difference in predicted outcomes under treatment versus control. More sophisticated versions fit separate models for treated and control groups, or use multi-stage procedures that directly target the conditional average treatment effect \citep{daoud2022gnf}. The choice among meta-learners involves tradeoffs: simpler approaches are more transparent but may be less efficient; complex approaches may achieve better statistical properties but at the cost of interpretability.

Both traditional and causal machine learning methods require the researcher to choose which covariates to include \citep{chernozhukov2024causalml}. For interaction models, this choice determines which effects can be estimated. For causal machine learning, this choice determines which sources of heterogeneity can be detected. In neither case does the method automatically discover heterogeneity in variables the researcher failed to measure. This point is worth emphasizing because causal machine learning methods are sometimes described as finding patterns ``in the data.'' More precisely, they find patterns in the variables the researcher chose to include, which may or may not capture the true sources of effect variation.

Table~\ref{tab:comparison} summarizes key features of these approaches across several dimensions relevant for practical method selection.

\begin{table}[htbp]
\centering
\caption{Comparing Traditional and Machine Learning Approaches to Effect Heterogeneity}
\label{tab:comparison}
\small
\begin{tabular}{p{3cm}p{5.5cm}p{5.5cm}}
\toprule
\textbf{Dimension} & \textbf{Interaction Models} & \textbf{Machine Learning} \\
\midrule
Logic & Deductive: test pre-specified hypotheses & Inductive: discover patterns from data \\
\addlinespace
Functional form & Specified by researcher (typically linear) & Learned from data (flexible) \\
\addlinespace
Interpretability & High: coefficients have direct meaning & Lower: requires additional analysis \\
\addlinespace
Sample requirements & Moderate for two-way; large for higher-order & Large (typically $n > 1000$) \\
\addlinespace
Multiple testing & Controlled if hypotheses pre-specified & Risk of overfitting spurious patterns \\
\addlinespace
High-dimensional covariates & Cannot handle e.g, large tabular data, text, images, or audio & Designed for high dimensions \\
\addlinespace
Best use case & Theory evaluation: Confirmatory analysis with strong theory & Theory development: Exploratory analysis with many potential moderators \\
\bottomrule
\end{tabular}
\end{table}

The practical choice between these approaches depends on research context. When theory provides strong guidance about which variables moderate effects and researchers seek to test these predictions, traditional interaction models are appropriate. When the goal is theorizing or exploratory---discovering unexpected patterns of heterogeneity without strong prior hypotheses---causal machine learning methods are valuable. When both discovery and confirmation matter, the two-stage approach combining inductive discovery with deductive follow-up offers a principled path forward.

A special challenge arises when covariates are high-dimensional---particularly when researchers wish to condition on images, text, or other complex data types. Standard causal machine learning methods like causal forests were designed for moderate-dimensional tabular data and struggle when applied to raw pixel values or word embeddings. Simply flattening an image into thousands of pixel intensities and feeding it to a causal forest will typically fail because the algorithm cannot learn meaningful visual representations from raw inputs. For such applications, researchers need specialized methods that combine computer vision or natural language processing with causal inference \citep{jerzak2023image, zhu2024multiscale}. These approaches typically learn low-dimensional representations $g(X)$ that capture causally relevant variation in complex inputs, then estimate heterogeneous effects conditional on these learned representations. This two-stage strategy---first dimensionality reduction, then causal estimation---provides a path toward meaningful conditioning in high-dimensional settings while acknowledging that the choice of representation matters substantively for which sources of heterogeneity can be detected.
\section{Simulation: When Does Each Approach Excel?}

This section presents a worked-example simulation. The results exemplify when traditional interaction models outperform machine learning and vice versa.

\subsection{Simulation Design}

We generated data with 2,000 observations and five covariates: three continuous variables representing income, test scores, and neighborhood quality, plus two binary variables representing minority status and gender. Treatment assignment varied with income (propensity score of 0.4 for below-median income, 0.6 for above-median), introducing realistic selection into treatment. We constructed three scenarios representing different underlying patterns of effect heterogeneity.

Figure~\ref{fig:simdag} displays the causal structure underlying our simulation. Income serves as a confounder with moderation: it affects both treatment assignment and the outcome, and it also modifies the treatment effect. Minority status and gender are pure moderators: they modify the treatment effect but do not influence treatment assignment in our design. This distinction matters because confounders must be adjusted for to identify causal effects, while pure moderators need not be---though both can legitimately appear in CATE conditioning sets. Neighborhood quality affects only the outcome (not treatment assignment) and does not modify effects, serving as a precision covariate. This structure illustrates that CATE estimation can condition on variables that play different roles in the causal graph.

\begin{figure}[htbp]
\centering
\begin{tikzpicture}[
    node distance=2cm,
    observed/.style={circle, draw=black, thick, minimum size=0.9cm, fill=white},
    arrow/.style={-{Stealth[scale=1.2]}, thick},
    modarrow/.style={-{Stealth[scale=1.2]}, thick, dashed, blue!70}
]
\node[observed] (W) {$W$};
\node[observed] (Y) [right=3cm of W] {$Y$};
\node[observed] (Inc) [above left=1.5cm and 1cm of W] {\small Inc};
\node[observed] (Min) [above=1.5cm of Y] {\small Min};
\node[observed] (Fem) [above right=1.5cm and 0.5cm of Y] {\small Fem};
\node[observed] (Nbh) [below=1.2cm of Y] {\small Nbh};
\node[observed, dashed] (U) [right=1.2cm of Y] {\small $U$};

\draw[arrow] (W) -- (Y);
\draw[arrow] (Inc) -- (W);
\draw[arrow] (Inc) to[bend left=15] (Y);
\draw[arrow] (Nbh) -- (Y);
\draw[arrow, dashed] (U) -- (Y);

\draw[modarrow] (Inc) to[out=-30, in=150] ($(W)!0.5!(Y)$);
\draw[modarrow] (Min) -- ($(W)!0.5!(Y)+(0,0.3)$);
\draw[modarrow] (Fem) to[out=200, in=40] ($(W)!0.5!(Y)+(0.2,0.2)$);

\node[below=0.3cm of Nbh, text width=8cm, align=center] {\footnotesize Solid arrows: causal effects. Dashed blue arrows: effect modification.\\Income is a confounder with moderation; Minority and Female are pure moderators.};
\end{tikzpicture}
\caption{Causal structure of the simulation. Income (Inc) is a confounder that also modifies the treatment effect. Minority status (Min) and gender (Fem) modify effects but do not affect treatment assignment---they are pure moderators. Neighborhood quality (Nbh) affects outcomes only and does not moderate effects. The unobserved noise term $U$ represents individual-specific factors that affect outcomes; recovering $U$ through abduction is essential for counterfactual inference at the individual level.} 
\label{fig:simdag}
\end{figure}

In the first scenario, the true conditional average treatment effect follows a simple linear interaction: the effect is 2 plus 1.5 times minority status. This scenario favors traditional methods because the data-generating process matches linear interaction assumptions. In the second scenario, the true effect follows a complex pattern combining discrete three-way interactions with continuous nonlinearity: the effect depends on the combination of minority status, gender, and whether income exceeds the median, plus a nonlinear function of test scores (specifically, the positive part of the test score). This scenario favors machine learning because no simple interaction model can capture both discrete and continuous sources of heterogeneity. In the third scenario, the true effect is constant at 2 for everyone, representing no heterogeneity.

For each scenario, we estimated conditional average treatment effects using two approaches: a fully saturated interaction model including all main effects and interactions, and a causal forest with 2,000 trees using the \texttt{grf} package in R, following best practices for ``honest'' estimation with separate subsamples for model building and effect estimation \citep{athey2019, shiba2024}. We evaluated performance using mean squared error---the average squared difference between estimated and true effects across all individuals.

\subsection{Results}

Table~\ref{tab:mse} presents mean squared error for each method and scenario.

\begin{table}[htbp]
\centering
\caption{Bias, Variance, and Mean Squared Error by Method and Scenario}
\label{tab:mse}
\small
\begin{tabular}{lcccccc}
\toprule
& \multicolumn{3}{c}{\textbf{OLS Interaction}} & \multicolumn{3}{c}{\textbf{Causal Forest}} \\
\cmidrule(lr){2-4} \cmidrule(lr){5-7}
\textbf{Scenario} & Bias & Variance & MSE & Bias & Variance & MSE \\
\midrule
Linear interaction & 0.062 & 0.014 & 0.018 & 0.033 & 0.011 & 0.012 \\
Complex nonlinear & 0.043 & 0.495 & 0.496 & 0.017 & 0.161 & 0.161 \\
Constant effect & 0.056 & 0.009 & 0.013 & 0.016 & 0.013 & 0.013 \\
\bottomrule
\end{tabular}

\vspace{0.5em}
\footnotesize
\textit{Note:} MSE $\approx$ Bias$^2$ + Variance. In the complex nonlinear scenario, OLS exhibits high variance because it cannot capture the true nonlinear heterogeneity pattern. The causal forest achieves lower MSE primarily through reduced variance.
\end{table}

The results illuminate when each approach excels. In the linear interaction scenario, both methods perform well, with the causal forest achieving slightly lower error despite the data-generating process matching the OLS specification. This occurs because the causal forest's regularization prevents overfitting to noise in the higher-order interactions that OLS includes but which have no true effect. In the complex nonlinear scenario, the causal forest dramatically outperforms OLS, achieving roughly one-third the error. The linear interaction model cannot capture the true pattern---which involves both discrete three-way interactions and nonlinear effects of continuous covariates---while the causal forest discovers these complexities automatically. In the constant effect scenario, both methods perform comparably, with OLS achieving marginally lower error. When no heterogeneity exists, the causal forest's flexibility becomes a slight liability, as the algorithm finds small amounts of spurious variation that OLS correctly ignores.

\begin{figure}[htbp]
\centering
\includegraphics[width=\textwidth]{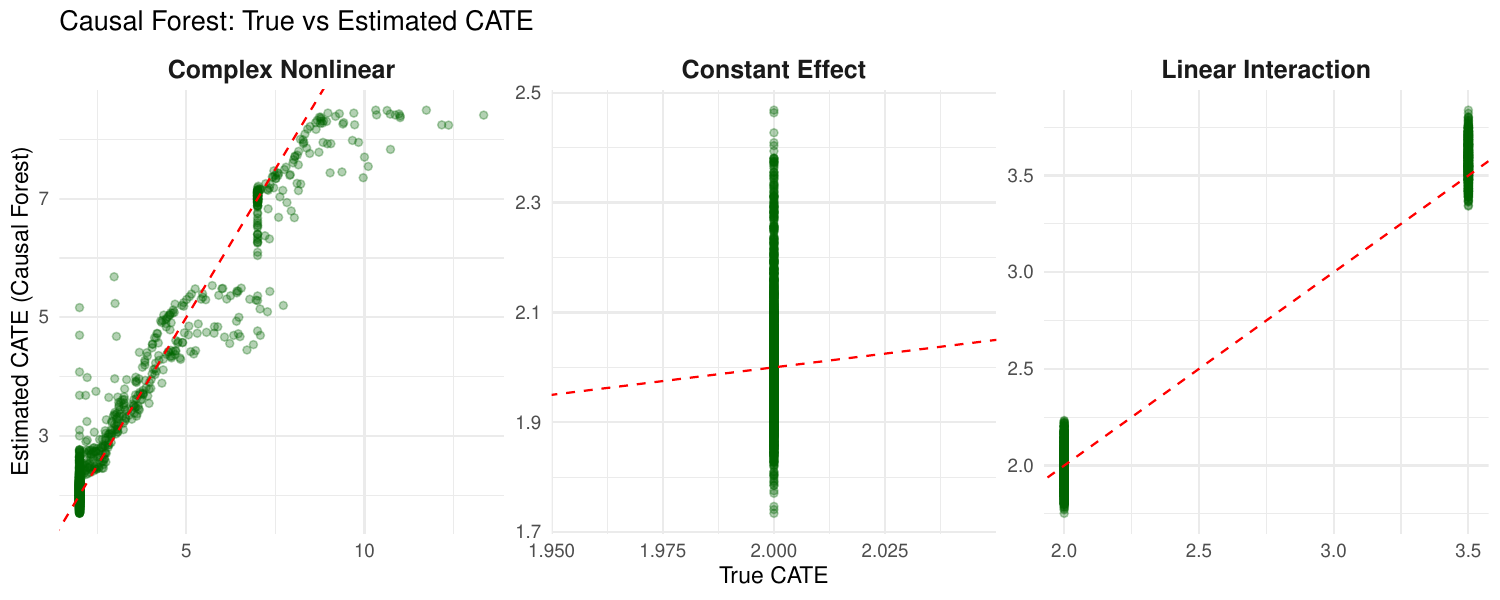}
\caption{Causal forest estimates versus true conditional average treatment effects across three simulation scenarios. Points falling along the diagonal indicate accurate estimation.}
\label{fig:scatter}
\end{figure}

Figure~\ref{fig:scatter} displays these patterns visually, plotting estimated against true effects for each method in the complex nonlinear scenario where differences are most pronounced.

\begin{figure}[htbp]
\centering
\includegraphics[width=0.6\textwidth]{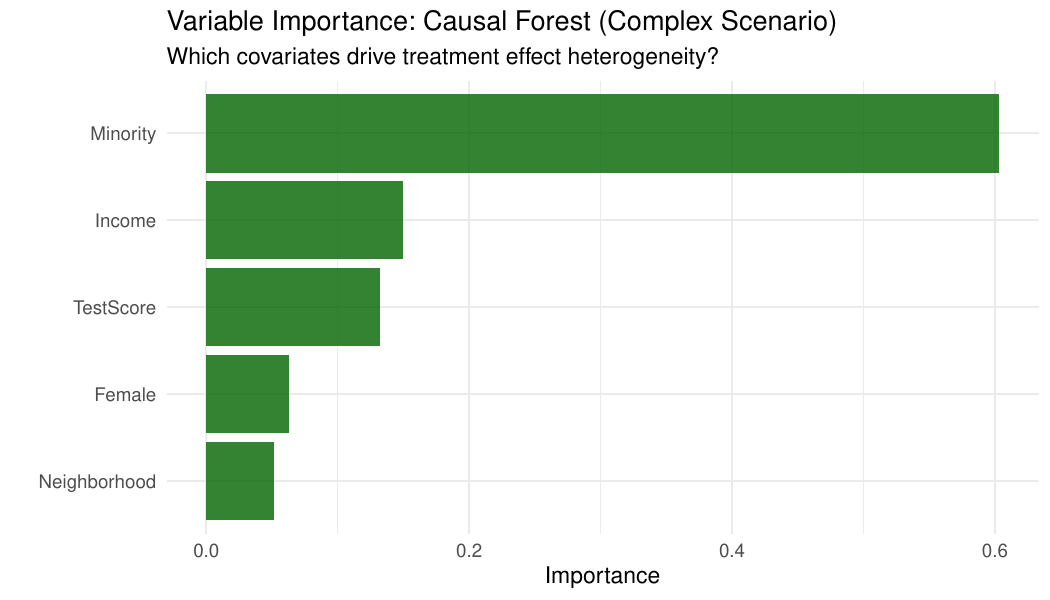}
\caption{Variable importance from causal forest in the complex nonlinear scenario, showing which covariates drive treatment effect heterogeneity.}
\label{fig:importance}
\end{figure}

The causal forest also provides information about which covariates drive heterogeneity. Figure~\ref{fig:importance} shows variable importance scores from the complex scenario, indicating that minority status and gender---the variables involved in the true interaction---receive the highest importance, while income (involved only through its median split) receives lower importance. This diagnostic can guide researchers toward interpretable summaries of machine learning results. A major advantage of these variable importance measures is that they evaluate the contribution of each variable across all covariates simultaneously, unlike interaction models which test only for pre-specified moderators. If a researcher omits an important moderator from an interaction model, they cannot detect its importance; causal forests screen all included covariates automatically. However, these importance metrics have a significant limitation: they report only marginal importance and do not reveal how variables interact with one another. A variable may receive high importance because it participates in a two-way interaction, a three-way interaction, or a complex nonlinear pattern---the importance score alone does not distinguish these cases. Researchers must conduct additional analyses, such as partial dependence plots or subgroup comparisons, to understand the structure of discovered heterogeneity.

\subsection{Subgroup-Specific Analysis: Targeting the True CATEs}

While aggregate metrics like MSE provide useful summaries, applied researchers ultimately care about recovering treatment effects for specific subgroups---the actual targets of heterogeneity analysis. To evaluate this directly, we examine how well each method estimates the true CATE for theoretically meaningful subgroups defined by the data-generating process.

In the linear interaction scenario, where the true effect is $\tau(x) = 2 + 1.5 \times \text{Minority}$, both methods accurately recover the subgroup-specific effects. Non-minority individuals have a true CATE of 2.0, and minority individuals have a true CATE of 3.5. OLS estimates these as 2.01 and 3.64 respectively, while the causal forest estimates 2.00 and 3.59. When the true heterogeneity follows the model's functional form, both approaches succeed.

The complex nonlinear scenario reveals a starkly different pattern. Here the true effect involves a three-way interaction: $\tau(x) = 2 + 5 \times \text{Minority} \times \text{Female} \times \mathbf{1}(\text{Income} > 0) + 2 \times \max(\text{TestScore}, 0) \times \text{Minority}$. Only one subgroup---minority women with above-median income---receives the additional +5 boost from the discrete interaction, while all minorities receive a contribution from the TestScore term. Table~\ref{tab:subgroup} presents the subgroup-specific results.

\begin{table}[htbp]
\centering
\caption{Subgroup-Specific CATE Estimates: Complex Nonlinear Scenario}
\label{tab:subgroup}
\small
\begin{tabular}{lcccccc}
\toprule
\textbf{Subgroup} & \textbf{N} & \textbf{True} & \textbf{OLS} & \textbf{CF} & \textbf{OLS Bias} & \textbf{CF Bias} \\
\midrule
Minority, Female, High Income & 188 & 7.83 & 5.90 & \textbf{7.29} & $-1.93$ & $\mathbf{-0.55}$ \\
Minority, Male, High Income & 225 & 2.91 & \textbf{3.06} & 3.13 & $\mathbf{+0.14}$ & $+0.22$ \\
Minority, Female, Low Income & 198 & 2.75 & 5.35 & \textbf{3.04} & $+2.60$ & $\mathbf{+0.29}$ \\
Non-Minority, Female, High Income & 264 & 2.00 & \textbf{2.07} & 1.93 & $\mathbf{+0.07}$ & $-0.07$ \\
Non-Minority, Male, High Income & 305 & 2.00 & 2.32 & \textbf{1.95} & $+0.32$ & $\mathbf{-0.05}$ \\
Non-Minority, Female, Low Income & 291 & 2.00 & 1.60 & \textbf{2.05} & $-0.40$ & $\mathbf{+0.05}$ \\
Minority, Male, Low Income & 203 & 2.75 & \textbf{2.54} & 2.78 & $\mathbf{-0.21}$ & $+0.03$ \\
Non-Minority, Male, Low Income & 326 & 2.00 & 1.79 & \textbf{2.09} & $-0.21$ & $\mathbf{+0.09}$ \\
\midrule
\textit{Mean Absolute Bias} & & & & & 0.74 & \textbf{0.17} \\
\bottomrule
\end{tabular}

\vspace{0.5em}
\footnotesize
\textit{Note:} True CATE for minorities includes contributions from both the discrete interaction (+5 for female high-income) and the continuous term ($2 \times \max(\text{TestScore}, 0)$). Bold values indicate the method with lower absolute bias for each subgroup. The causal forest achieves lower bias in 6 of 8 subgroups. The key 3-way interaction subgroup (Minority, Female, High Income) receives a +5 boost that OLS substantially underestimates.
\end{table}

The key finding emerges in the first row: minority women with high income have a true CATE of 7.83, but OLS estimates only 5.90---underestimating by 25\%. Because the OLS model does not include the three-way interaction term, it cannot detect this unique concentration of treatment effects. The causal forest, which discovered this pattern inductively, estimates 7.29---substantially closer to the truth.

Equally telling is the pattern for minority women with \textit{low} income. This subgroup has a true CATE of only 2.75 (no interaction boost, but some TestScore contribution), yet OLS estimates 5.35---\textit{overestimating} by 95\%. Without the ability to distinguish between high and low income minority women, OLS spreads the interaction effect across all minority women regardless of income, biasing estimates in opposite directions for the two income groups. The causal forest correctly recognizes that low-income minority women do not receive the interaction boost, estimating 3.04.

\begin{figure}[htbp]
\centering
\includegraphics[width=\textwidth]{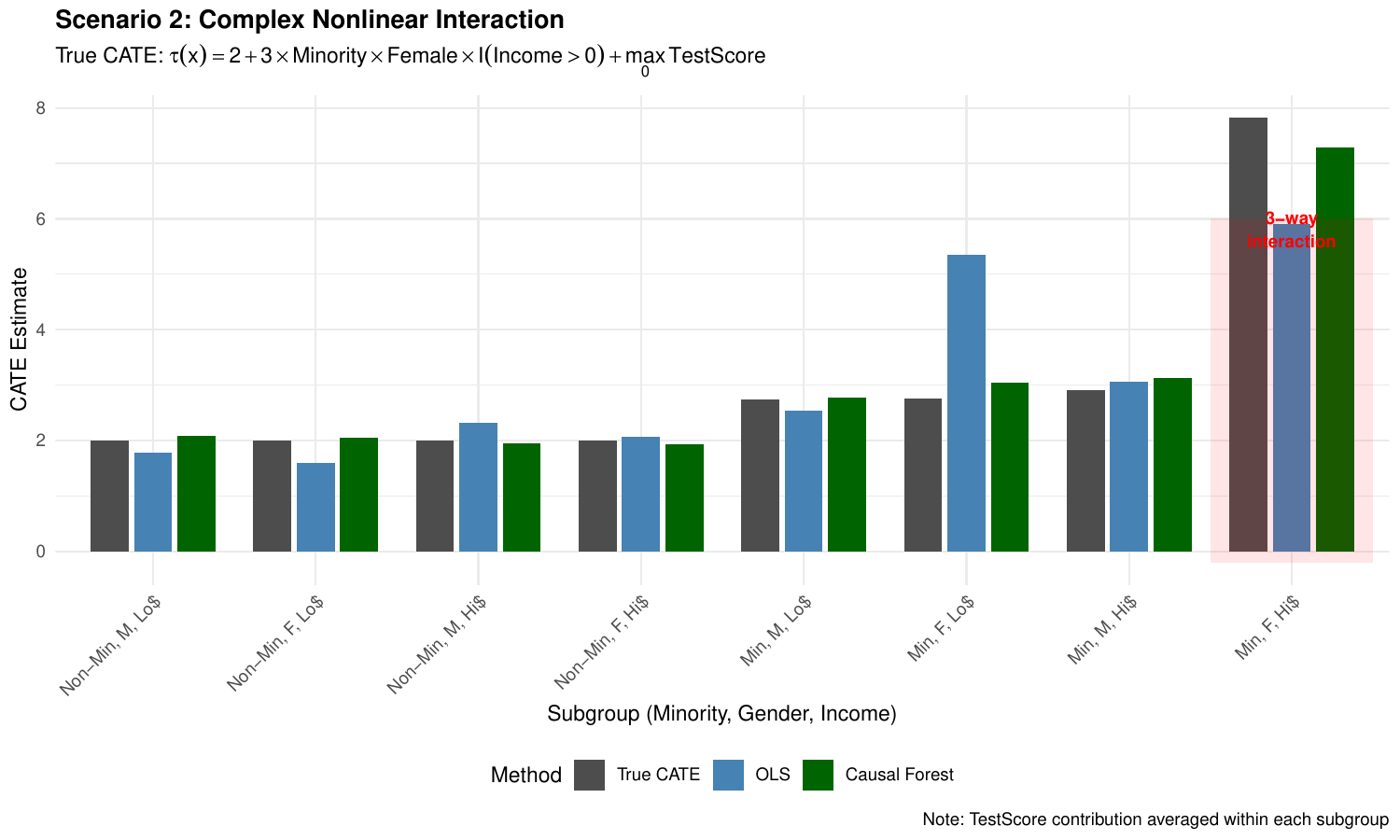}
\caption{True versus estimated CATE by subgroup in the complex nonlinear scenario. The causal forest (green) tracks true effects (gray) more closely than OLS (blue) across most subgroups, with the largest discrepancy for minority women where OLS cannot distinguish between high and low income (see rows 1 and 3 of Table~\ref{tab:subgroup}). The causal forest achieves over four times lower mean absolute bias than OLS (0.17 vs.\ 0.74).}
\label{fig:subgroup}
\end{figure}

Figure~\ref{fig:subgroup} visualizes these patterns. Across all eight subgroups, the causal forest's estimates (green bars) track the true effects (gray bars) more closely than OLS (blue bars). The mean absolute bias across subgroups is 0.17 for the causal forest versus 0.74 for OLS---the causal forest achieves over four times the accuracy at the subgroup level.

This analysis demonstrates why the choice between deductive and inductive approaches matters substantively, not just statistically. If researchers sought to identify which subgroup benefits most from treatment---a core goal of heterogeneity analysis---OLS would point them toward minority women generally, while the causal forest would correctly identify minority women with above-median income as the distinctive beneficiaries. For policy targeting, this distinction could mean the difference between effective and wasteful intervention allocation.

\subsection{What About Individual Treatment Effects?}
The methods discussed so far---interaction models and causal forests---estimate conditional average treatment effects: the expected effect for individuals sharing certain covariate values. But sometimes researchers seek something more ambitious: the individual treatment effect (ITE) for a specific person. What would \emph{this} person's outcome have been under the counterfactual treatment? This question requires moving beyond reduced-form estimation to full counterfactual inference \citep{pearl2009, pearl2009counterfactual}.

Pearl's framework distinguishes three steps in counterfactual reasoning \citep{pearl2009counterfactual}. First, \emph{abduction}: given the observed evidence (this person's covariates, treatment, and outcome), infer the latent factors that produced these observations. Second, \emph{action}: modify the causal model to reflect the counterfactual intervention (e.g., setting treatment to the opposite value). Third, \emph{prediction}: compute what the outcome would have been under this modified model. This three-step procedure requires knowing not just conditional independence relationships (what standard causal inference assumes) but the full structural equations governing each variable.

When structural forms are unknown, they can be learned from data using methods like Causal Graphical Normalizing Flows (cGNF) \citep{balgi2022cgnf, daoud2022gnf, balgi2024deeplearning}. These methods model the joint distribution of all variables in the DAG, enabling researchers to perform counterfactual inference at the individual level. Recent advances extend this framework to handle instrumental variables and more complex identification strategies \citep{braun2025flowiv}. Importantly, causal forests and other reduced-form methods \emph{cannot} estimate true individual treatment effects because they only estimate expectations conditional on covariates---they do not model the full structural equations needed for counterfactual queries. Recent work extends these methods to handle unobserved confounding through sensitivity analysis \citep{balgi2025sensitivity, lin2023sensitivity, balgi2022rho}.

To illustrate, consider two individuals from our simulation. Individual 1 is a minority woman with above-median income and above-median test scores---the subgroup where our data-generating process assigns the highest treatment effects. Individual 2 is a non-minority man with below-median income and below-median test scores---expected to have a minimal effect. For Individual 1, the true ITE is 6.0, comprising the base effect (2), the three-way interaction bonus (3), and a contribution from test scores. For Individual 2, the true ITE is exactly 2.0---the base effect with no interaction boost. These individual-level effects differ from the subgroup CATEs (5.3 and 2.4, respectively) because each person's outcome reflects their unique combination of characteristics, not just the subgroup average.

\begin{figure}[h]
\centering
\includegraphics[width=0.7\textwidth]{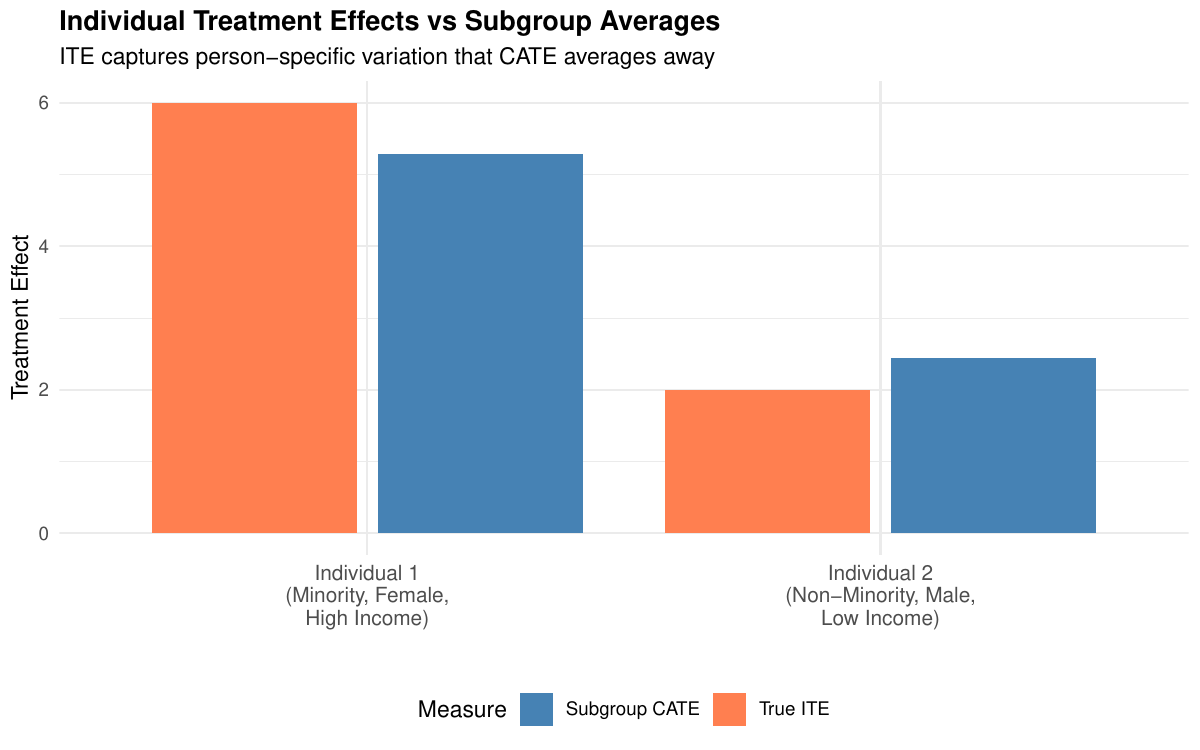}
\caption{Individual treatment effects versus subgroup conditional average treatment effects. The ITE captures person-specific variation that CATE averages away. For Individual 1 (minority, female, high income), the true ITE (6.0) exceeds the subgroup CATE (5.3) due to their specific covariate values. For Individual 2 (non-minority, male, low income), the true ITE (2.0) falls below the subgroup CATE (2.4).}
\label{fig:ite}
\end{figure}

The simulation also illuminates the relationship between conditional and individual treatment effects. Even with perfect knowledge of the data-generating process, uncertainty remains at the individual level. The causal forest provides not only point estimates but also standard errors for each individual's effect. In the complex scenario, 95 percent confidence intervals for individual effects typically span about 0.5 units---substantial uncertainty even when the method correctly captures average patterns within subgroups.

This finding has important implications for policy targeting. When decision-makers seek to identify individuals who would benefit most from treatment, they must reckon with this irreducible uncertainty. A person with an estimated effect of 0.7 and another with an estimated effect of 0.5 may not truly differ---their confidence intervals overlap substantially. Machine learning enables fine-grained estimation, but it cannot eliminate the fundamental problem that individual treatment effects are unobservable.

\section{Discussion}

The preceding analysis demonstrates that researchers studying effect heterogeneity face a fundamental methodological choice: whether to approach the problem deductively, specifying theoretical hypotheses about which variables moderate effects, or inductively, allowing algorithms to discover patterns in data. Our simulation results show that neither approach dominates uniformly---each excels in different contexts. When true heterogeneity follows simple patterns that match linear interaction assumptions, traditional OLS models perform well. When heterogeneity involves complex higher-order interactions or nonlinearities, causal machine learning methods like causal forests substantially outperform pre-specified models. This finding carries direct implications for applied research: the choice of method should reflect both the researcher's theoretical priors and the likely complexity of underlying heterogeneity patterns.

Effect heterogeneity methods speak directly to core questions in the sociology of inequality and beyond. This section develops three connections, drawing on recent applications in domains ranging from educational stratification to global health \citep{brand2010, xiebrandjann2012, daoud2019grf}.

Researchers increasingly seek to understand not just the average effect of an intervention but how it would change existing disparities. The gap-closing estimand \citep{lundberg2021} formalizes this question: how much would the gap between two groups (say, racial income differences) close if we equalized a treatment (say, college access)? This estimand reframes conditional average treatment effect estimation as directly policy-relevant. Studies of international policy interventions illustrate this approach: average effects of structural adjustment programs on child health mask substantial variation across country contexts, with some nations experiencing protective effects while others face harm \citep{daoud2019imf, daoud2019ije}. If the return to college is higher for White students than for Black students, expanding college access alone will not close racial income gaps---and might even widen them. Effect heterogeneity analysis transforms abstract causal inference into concrete guidance for anti-inequality interventions.

Intersectionality theory emphasizes that social positions are defined by multiple, interacting dimensions of stratification \citep{crenshaw1989, mccall2005, bauer2021}. A Black woman does not experience the sum of disadvantages facing Black people and women separately; her social position is qualitatively distinct from either group considered alone. Effect heterogeneity methods can operationalize this theoretical insight by estimating how treatment effects vary across combinations of race, class, and gender \citep{daoud2021heterogeneous}. Traditional interaction models struggle here because the number of possible interactions grows combinatorially with the number of dimensions. Machine learning methods can navigate this high-dimensional space, discovering subgroups where effects concentrate without requiring researchers to specify all interactions in advance \citep{daoud2024scoping, daoud2022image}. The simulation results demonstrate this advantage: when true heterogeneity involves complex interactions, causal forests substantially outperform linear models.

At the same time, researchers must exercise caution when applying these methods to inequality questions \citep{daoud2019maternal, daoud2017}. Targeting interventions toward individuals with the highest estimated treatment effects---a natural application of fine-grained effect estimation---can reinforce existing inequalities if those individuals are systematically advantaged on other dimensions. Machine learning algorithms optimize statistical criteria that may not align with normative goals of equity and fairness \citep{daoud2023debiasing}. Effect heterogeneity research in the inequality domain must engage with these normative questions, not merely technical optimization.

\subsection{Recommendations for Practice}

For applied researchers, we offer three recommendations. First, specify estimands before choosing methods. The research question should determine whether average effects, subgroup effects, or individual-level predictions are the appropriate target. Second, match methods to research goals. Use interaction models for confirmatory tests of theoretically motivated hypotheses; use causal machine learning for exploratory discovery; use two-stage approaches when both matter \citep{daoud2019grf, daoud2021melting, daoud2024scoping}. Third, report methods transparently. For causal machine learning results, describe variable importance, sensitivity to tuning parameters, and validation procedures. For interaction models, follow best practices for presenting and interpreting interaction effects \citep{brambor2006}.

Effect heterogeneity is not a nuisance to be controlled but a phenomenon central to sociological inquiry. Understanding how causal effects vary across populations illuminates the mechanisms of inequality, informs policy targeting, and builds richer theories of social processes. By clarifying the conceptual foundations and providing practical guidance for method selection, we hope to equip researchers to pursue this investigation with both rigor and relevance.



\appendix
\section{Formal Definitions}
\label{app:definitions}

This appendix provides formal definitions of the estimands discussed in the main text.

\begin{definition}[Average Treatment Effect]
For a binary treatment $W \in \{0,1\}$ and potential outcomes $Y(1)$ and $Y(0)$, the \textbf{average treatment effect (ATE)} is:
\begin{equation}
\tau = \mathbb{E}[Y(1) - Y(0)]
\end{equation}
\end{definition}

\begin{definition}[Conditional Average Treatment Effect]
For covariates $X$, the \textbf{conditional average treatment effect (CATE)} is:
\begin{equation}
\tau(x) = \mathbb{E}[Y(1) - Y(0) \mid X = x]
\end{equation}
\end{definition}

\begin{definition}[Individual Treatment Effect]
For unit $i$, the \textbf{individual treatment effect (ITE)} is:
\begin{equation}
\tau_i = Y_i(1) - Y_i(0)
\end{equation}
\end{definition}

\subsection{Identification Assumptions}

Causal identification requires the following assumptions:

\begin{assumption}[Conditional Ignorability]
Treatment assignment is independent of potential outcomes conditional on observed covariates:
\begin{equation}
\{Y(0), Y(1)\} \perp\!\!\!\perp W \mid X
\end{equation}
\end{assumption}

\begin{assumption}[Overlap/Positivity]
All covariate values have positive probability of each treatment:
\begin{equation}
0 < P(W = 1 \mid X = x) < 1 \quad \text{for all } x
\end{equation}
\end{assumption}

\begin{assumption}[SUTVA]
The stable unit treatment value assumption requires no interference between units and no hidden treatment variations.
\end{assumption}

\section{Meta-Learners for CATE Estimation}
\label{app:metalearners}

Meta-learners transform CATE estimation into standard prediction problems. Let $\hat{\mu}(w, x)$ denote a model predicting $Y$ given treatment $w$ and covariates $x$.

\textbf{S-learner.} Fit a single model $\hat{\mu}(W, X)$ and compute:
\begin{equation}
\hat{\tau}(x) = \hat{\mu}(1, x) - \hat{\mu}(0, x)
\end{equation}

\textbf{T-learner.} Fit separate models $\hat{\mu}_1(x)$ for treated and $\hat{\mu}_0(x)$ for control:
\begin{equation}
\hat{\tau}(x) = \hat{\mu}_1(x) - \hat{\mu}_0(x)
\end{equation}

\textbf{X-learner.} A two-stage approach that imputes treatment effects and models them as a function of covariates, with weighting by propensity scores.

\textbf{R-learner.} Directly targets CATE by minimizing:
\begin{equation}
\hat{\tau} = \arg\min_\tau \sum_i \left( Y_i - \hat{m}(X_i) - (W_i - \hat{e}(X_i))\tau(X_i) \right)^2
\end{equation}
where $\hat{m}$ and $\hat{e}$ are outcome and propensity models.

\textbf{DR-learner.} Uses doubly robust scores for efficiency:
\begin{equation}
\hat{\Gamma}_i = \hat{\mu}_1(X_i) - \hat{\mu}_0(X_i) + \frac{W_i(Y_i - \hat{\mu}_1(X_i))}{\hat{e}(X_i)} - \frac{(1-W_i)(Y_i - \hat{\mu}_0(X_i))}{1-\hat{e}(X_i)}
\end{equation}

\section{Simulation Code}
\label{app:code}

Full R code for replicating our simulation is available in the supplementary materials file \texttt{simulation.R}. The code uses the \texttt{grf} package for causal forests and base R for interaction models.


\newpage
\bibliographystyle{apalike}
\bibliography{references}

@book{morganwinship2015,
  author = {Morgan, Stephen L. and Winship, Christopher},
  title = {Counterfactuals and Causal Inference: Methods and Principles for Social Research},
  edition = {2nd},
  publisher = {Cambridge University Press},
  year = {2015}
}

@book{pearl2009,
  author = {Pearl, Judea},
  title = {Causality: Models, Reasoning, and Inference},
  edition = {2nd},
  publisher = {Cambridge University Press},
  year = {2009}
}

@book{aikenwest1991,
  author = {Aiken, Leona S. and West, Stephen G.},
  title = {Multiple Regression: Testing and Interpreting Interactions},
  publisher = {Sage},
  year = {1991}
}

@article{brambor2006,
  author = {Brambor, Thomas and Clark, William Roberts and Golder, Matt},
  title = {Understanding Interaction Models: Improving Empirical Analyses},
  journal = {Political Analysis},
  year = {2006},
  volume = {14},
  number = {1},
  pages = {63--82}
}

@article{hainmueller2019,
  author = {Hainmueller, Jens and Mummolo, Jonathan and Xu, Yiqing},
  title = {How Much Should We Trust Estimates from Multiplicative Interaction Models? Simple Tools to Improve Empirical Practice},
  journal = {Political Analysis},
  year = {2019},
  volume = {27},
  number = {2},
  pages = {163--192}
}

@misc{gelman2018,
  author = {Gelman, Andrew},
  title = {You Need 16 Times the Sample Size to Estimate an Interaction Than to Estimate a Main Effect},
  year = {2018},
  howpublished = {Statistical Modeling, Causal Inference, and Social Science (blog)},
  url = {https://statmodeling.stat.columbia.edu/2018/03/15/need-16-times-sample-size-estimate-interaction-estimate-main-effect/}
}

@article{atheywager2018,
  author = {Athey, Susan and Wager, Stefan},
  title = {Estimation and Inference of Heterogeneous Treatment Effects using Random Forests},
  journal = {Journal of the American Statistical Association},
  year = {2018},
  volume = {113},
  number = {523},
  pages = {1228--1242}
}

@article{athey2019,
  author = {Athey, Susan and Tibshirani, Julie and Wager, Stefan},
  title = {Generalized Random Forests},
  journal = {Annals of Statistics},
  year = {2019},
  volume = {47},
  number = {2},
  pages = {1148--1178}
}

@article{kunzel2019,
  author = {K\"{u}nzel, S\"{o}ren R. and Sekhon, Jasjeet S. and Bickel, Peter J. and Yu, Bin},
  title = {Metalearners for Estimating Heterogeneous Treatment Effects Using Machine Learning},
  journal = {Proceedings of the National Academy of Sciences},
  year = {2019},
  volume = {116},
  number = {10},
  pages = {4156--4165}
}

@article{nie2021,
  author = {Nie, Xinkun and Wager, Stefan},
  title = {Quasi-Oracle Estimation of Heterogeneous Treatment Effects},
  journal = {Biometrika},
  year = {2021},
  volume = {108},
  number = {2},
  pages = {299--319}
}

@article{kennedy2023,
  author = {Kennedy, Edward H.},
  title = {Towards Optimal Doubly Robust Estimation of Heterogeneous Causal Effects},
  journal = {Electronic Journal of Statistics},
  year = {2023},
  volume = {17},
  number = {2},
  pages = {3008--3049}
}

@article{brand2023,
  author = {Brand, Jennie E. and Zhou, Xiaoping and Xie, Yu},
  title = {Recent Developments in Causal Inference and Machine Learning},
  journal = {Annual Review of Sociology},
  year = {2023},
  volume = {49},
  pages = {81--110}
}

@article{molinagarip2019,
  author = {Molina, Mario and Garip, Filiz},
  title = {Machine Learning for Sociology},
  journal = {Annual Review of Sociology},
  year = {2019},
  volume = {45},
  pages = {27--45}
}

@article{lundberg2021,
  author = {Lundberg, Ian},
  title = {The Gap-Closing Estimand: A Causal Approach to Study Interventions That Close Disparities Across Social Categories},
  journal = {Sociological Methods \& Research},
  year = {2024},
  volume = {53},
  number = {2},
  pages = {507--570},
  doi = {10.1177/00491241211055769}
}

@article{lundberg2021what,
  author = {Lundberg, Ian and Johnson, Rebecca and Stewart, Brandon M.},
  title = {What Is Your Estimand? Defining the Target Quantity Connects Statistical Evidence to Theory},
  journal = {American Sociological Review},
  year = {2021},
  volume = {86},
  number = {3},
  pages = {532--565}
}

@article{brand2010,
  author = {Brand, Jennie E. and Xie, Yu},
  title = {Who Benefits Most from College? Evidence for Negative Selection in Heterogeneous Economic Returns to Higher Education},
  journal = {American Sociological Review},
  year = {2010},
  volume = {75},
  number = {2},
  pages = {273--302}
}

@article{mccall2005,
  author = {McCall, Leslie},
  title = {The Complexity of Intersectionality},
  journal = {Signs: Journal of Women in Culture and Society},
  year = {2005},
  volume = {30},
  number = {3},
  pages = {1771--1800}
}

@article{crenshaw1989,
  author = {Crenshaw, Kimberl\'{e}},
  title = {Demarginalizing the Intersection of Race and Sex: A Black Feminist Critique of Antidiscrimination Doctrine, Feminist Theory and Antiracist Politics},
  journal = {University of Chicago Legal Forum},
  year = {1989},
  volume = {1989},
  number = {1},
  pages = {139--167}
}

@article{bauer2021,
  author = {Bauer, Greta R. and Churchill, Siobhan M. and Mahendran, Mayuri and Walwyn, Chantel and Lizotte, Daniel and Villa-Rueda, Alma Angelica},
  title = {Intersectionality in Quantitative Research: A Systematic Review of Its Emergence and Applications of Theory and Methods},
  journal = {SSM - Population Health},
  year = {2021},
  volume = {14},
  pages = {100798}
}

@article{shiba2024,
  author = {Shiba, Koichiro and Inoue, Kosuke},
  title = {Harnessing Causal Forests for Epidemiologic Research: Key Considerations},
  journal = {American Journal of Epidemiology},
  year = {2024},
  note = {Commentary}
}

@article{holland1986,
  author = {Holland, Paul W.},
  title = {Statistics and Causal Inference},
  journal = {Journal of the American Statistical Association},
  year = {1986},
  volume = {81},
  number = {396},
  pages = {945--960}
}

@article{vegetabile2021,
  author = {Vegetabile, Brian G.},
  title = {On the Distinction Between ``Conditional Average Treatment Effects'' (CATE) and ``Individual Treatment Effects'' (ITE) Under Ignorability Assumptions},
  journal = {arXiv preprint arXiv:2108.04939},
  year = {2021},
  url = {https://arxiv.org/abs/2108.04939}
}

@article{vanderweele2009interaction,
  author = {VanderWeele, Tyler J. and Robins, James M.},
  title = {The Identification of Synergism in the Sufficient-Component-Cause Framework},
  journal = {Epidemiology},
  year = {2007},
  volume = {18},
  number = {3},
  pages = {329--339},
  doi = {10.1097/01.ede.0000260218.66432.88}
}

@article{vanderweele2009dag,
  author = {VanderWeele, Tyler J. and Robins, James M.},
  title = {Directed Acyclic Graphs, Sufficient Causes, and the Properties of Conditioning on a Common Effect},
  journal = {American Journal of Epidemiology},
  year = {2007},
  volume = {166},
  number = {9},
  pages = {1096--1104},
  doi = {10.1093/aje/kwm179}
}

@article{nilsson2021,
  author = {Nilsson, Anton and Bonander, Carl and Str\"{o}mberg, Ulf and Bj\"{o}rk, Jonas},
  title = {A Directed Acyclic Graph for Interactions},
  journal = {International Journal of Epidemiology},
  year = {2021},
  volume = {50},
  number = {2},
  pages = {613--619}
}

@book{hernan2020,
  author = {Hern\'{a}n, Miguel A. and Robins, James M.},
  title = {Causal Inference: What If},
  publisher = {Chapman \& Hall/CRC},
  year = {2020}
}

@article{hill2011,
  author = {Hill, Jennifer L.},
  title = {Bayesian Nonparametric Modeling for Causal Inference},
  journal = {Journal of Computational and Graphical Statistics},
  year = {2011},
  volume = {20},
  number = {1},
  pages = {217--240}
}

@article{xie2013,
  author = {Xie, Yu},
  title = {Population Heterogeneity and Causal Inference},
  journal = {Proceedings of the National Academy of Sciences},
  year = {2013},
  volume = {110},
  number = {16},
  pages = {6262--6268}
}

@article{xiebrandjann2012,
  author = {Xie, Yu and Brand, Jennie E. and Jann, Ben},
  title = {Estimating Heterogeneous Treatment Effects with Observational Data},
  journal = {Sociological Methodology},
  year = {2012},
  volume = {42},
  number = {1},
  pages = {314--347}
}

@article{brand2021,
  author = {Brand, Jennie E. and Xu, Jun and Koch, Bernard and Geraldo, Pablo},
  title = {Uncovering Sociological Effect Heterogeneity Using Tree-Based Machine Learning},
  journal = {Sociological Methodology},
  year = {2021},
  volume = {51},
  number = {1},
  pages = {189--223}
}

@article{daoud2017,
  author = {Daoud, Adel and Reinsberg, Bernhard and Kentikelenis, Alexander E. and Stubbs, Thomas H. and King, Lawrence P.},
  title = {The International Monetary Fund's Interventions in Food and Agriculture: An Analysis of Loans and Conditions},
  journal = {Food Policy},
  year = {2019},
  volume = {83},
  pages = {204--218},
  doi = {10.1016/j.foodpol.2019.01.005}
}

@article{daoud2019imf,
  author = {Daoud, Adel and Nosrati, Elias and Reinsberg, Bernhard and Kentikelenis, Alexander E. and Stubbs, Thomas H. and King, Lawrence P.},
  title = {Impact of International Monetary Fund Programs on Child Health},
  journal = {Proceedings of the National Academy of Sciences},
  year = {2017},
  volume = {114},
  number = {25},
  pages = {6492--6497}
}

@article{daoud2019grf,
  author = {Daoud, Adel and Johansson, Fredrik D.},
  title = {The Impact of Austerity on Children: {U}ncovering Effect Heterogeneity by Political, Economic, and Family Factors in Low- and Middle-Income Countries},
  journal = {Social Science Research},
  year = {2024},
  volume = {118},
  pages = {102973},
  doi = {10.1016/j.ssresearch.2023.102973}
}

@article{daoud2019ije,
  author = {Daoud, Adel and Reinsberg, Bernhard},
  title = {Structural Adjustment, State Capacity and Child Health: Evidence from {IMF} Programmes},
  journal = {International Journal of Epidemiology},
  year = {2019},
  volume = {48},
  number = {2},
  pages = {445--454},
  doi = {10.1093/ije/dyy251}
}

@article{daoud2019maternal,
  author = {Kraamwinkel, Nadine and Ekbrand, Hans and Davia, Stefania and Daoud, Adel},
  title = {The Influence of Maternal Agency on Severe Child Undernutrition in Conflict-Ridden Nigeria: {M}odeling Heterogeneous Treatment Effects with Machine Learning},
  journal = {PLOS ONE},
  year = {2019},
  volume = {14},
  number = {1},
  pages = {e0208937},
  doi = {10.1371/journal.pone.0208937}
}

@article{daoud2021heterogeneous,
  author = {Shiba, Koichiro and Daoud, Adel and Kino, Shiho and Nishi, Daisuke and Kondo, Katsunori and Kawachi, Ichiro},
  title = {Uncovering Heterogeneous Associations of Disaster-Related Traumatic Experiences with Subsequent Mental Health Problems: A Machine Learning Approach},
  journal = {Psychiatry and Clinical Neurosciences},
  year = {2022},
  volume = {76},
  number = {4},
  pages = {97--105},
  doi = {10.1111/pcn.13322}
}

@article{daoud2022image,
  author = {Jerzak, Connor T. and Johansson, Fredrik and Daoud, Adel},
  title = {Estimating Causal Effects Under Image Confounding Bias with an Application to Poverty in Africa},
  journal = {arXiv preprint arXiv:2206.06410},
  year = {2022}
}

@inproceedings{daoud2023debiasing,
  author = {Audinet de Pieuchon, Nicolas and Daoud, Adel and Jerzak, Connor T. and Johansson, Moa and Johansson, Richard},
  title = {Benchmarking Debiasing Methods for {LLM}-based Parameter Estimates},
  booktitle = {Proceedings of the 2025 Conference on Empirical Methods in Natural Language Processing (EMNLP)},
  year = {2025},
  pages = {17856--17870},
  publisher = {Association for Computational Linguistics},
  doi = {10.18653/v1/2025.emnlp-main.1000}
}

@incollection{daoud2024scoping,
  author = {Sakamoto, Kazuki and Jerzak, Connor T. and Daoud, Adel},
  title = {A Scoping Review of Earth Observation and Machine Learning for Causal Inference: Implications for the Geography of Poverty},
  booktitle = {Geography of Poverty},
  editor = {Hall, Ola and Wahab, Ibrahim},
  publisher = {Edward Elgar Publishing},
  address = {Cheltenham, UK},
  year = {2025},
  note = {arXiv:2406.02584}
}

@article{daoud2021melting,
  author = {Daoud, Adel and Dubhashi, Devdatt},
  title = {Melting Together Prediction and Inference},
  journal = {Observational Studies},
  year = {2021},
  volume = {7},
  number = {1},
  pages = {1--7}
}

@article{daoud2023hdsr,
  author = {Daoud, Adel and Dubhashi, Devdatt},
  title = {Statistical Modeling: The Three Cultures},
  journal = {Harvard Data Science Review},
  year = {2023},
  volume = {5},
  number = {1},
  doi = {10.1162/99608f92.89f6fe66}
}

@article{daoud2015poverty,
  author = {Daoud, Adel},
  title = {Quality of Governance, Corruption and Absolute Child Poverty in India},
  journal = {Journal of South Asian Development},
  year = {2015},
  volume = {10},
  number = {2},
  pages = {148--167},
  doi = {10.1177/0973174115588844}
}

@article{daoud2016satpov,
  author = {Daoud, Adel and Haller\"od, Bj\"orn and Guha-Sapir, Debarati},
  title = {What Is the Association between Absolute Child Poverty, Poor Governance, and Natural Disasters? {A} Global Comparison of Some of the Realities of Climate Change},
  journal = {PLOS ONE},
  year = {2016},
  volume = {11},
  number = {4},
  pages = {e0153296},
  doi = {10.1371/journal.pone.0153296}
}

@article{daoud2022gnf,
  author = {Balgi, Sourabh and Peña, Jose M. and Daoud, Adel},
  title = {Counterfactual Analysis of the Impact of the {IMF} Program on Child Poverty in the {Global-South} Region using Causal-Graphical Normalizing Flows},
  journal = {arXiv preprint arXiv:2202.09391},
  year = {2022}
}

@article{lundberg2025gapclosing,
  author = {Lundberg, Ian},
  title = {The Causal Impact of Segregation on a Disparity: A Gap-Closing Approach},
  journal = {Sociological Science},
  year = {2025},
  volume = {12},
  pages = {871--890},
  doi = {10.15195/v12.a35}
}

@article{mood2010logistic,
  author = {Mood, Carina},
  title = {Logistic Regression: Why We Cannot Do What We Think We Can Do, and What We Can Do About It},
  journal = {European Sociological Review},
  year = {2010},
  volume = {26},
  number = {1},
  pages = {67--82},
  doi = {10.1093/esr/jcp006}
}

@book{chernozhukov2024causalml,
  author = {Chernozhukov, Victor and Hansen, Christian and Kallus, Nathan and Spindler, Martin and Syrgkanis, Vasilis},
  title = {Applied Causal Inference Powered by {ML} and {AI}},
  publisher = {CausalML-book.org},
  year = {2024},
  note = {arXiv:2403.02467}
}

@inproceedings{jerzak2023image,
  author = {Jerzak, Connor T. and Johansson, Fredrik and Daoud, Adel},
  title = {Image-based Treatment Effect Heterogeneity},
  booktitle = {Proceedings of the Second Conference on Causal Learning and Reasoning (CLeaR)},
  series = {Proceedings of Machine Learning Research},
  volume = {213},
  pages = {1--22},
  year = {2023},
  publisher = {PMLR}
}

@article{zhu2024multiscale,
  author = {Zhu, Fucheng Warren and Jerzak, Connor T. and Daoud, Adel},
  title = {Optimizing Multi-Scale Representations to Detect Effect Heterogeneity Using Earth Observation and Computer Vision: Applications to Two Anti-Poverty {RCT}s},
  journal = {arXiv preprint arXiv:2411.02134},
  year = {2024}
}

@article{braun2025flowiv,
  author = {Braun, Felix and Wei{\ss}, Christoph and Gerwinn, Sebastian and Ortega, Pedro A.},
  title = {Flow {IV}: Counterfactual Inference with Flows for Instrumental Variables},
  journal = {arXiv preprint arXiv:2508.01321},
  year = {2025}
}

@article{balgi2022cgnf,
  author = {Balgi, Sourabh and Pe{\~n}a, Jose M. and Daoud, Adel},
  title = {Personalized Public Policy Analysis in Social Sciences Using Causal-Graphical Normalizing Flows},
  journal = {Proceedings of the AAAI Conference on Artificial Intelligence},
  year = {2022},
  volume = {36},
  number = {11},
  pages = {11810--11818}
}

@article{pearl2009counterfactual,
  author = {Pearl, Judea},
  title = {Causal Inference in Statistics: An Overview},
  journal = {Statistics Surveys},
  year = {2009},
  volume = {3},
  pages = {96--146}
}

@article{balgi2025sensitivity,
  author = {Balgi, Sourabh and Braun, Marc and Pe{\~n}a, Jos{\'e} M. and Daoud, Adel},
  title = {Sensitivity Analysis to Unobserved Confounding with Copula-based Normalizing Flows},
  journal = {International Journal of Approximate Reasoning},
  year = {2025},
  doi = {10.1016/j.ijar.2025.109391}
}

@article{lin2023sensitivity,
  author = {Lin, Cheng and Pe{\~n}a, Jos{\'e} M. and Daoud, Adel},
  title = {Navigating Unmeasured Confounding in Quantitative Sociology: A Sensitivity Framework},
  journal = {arXiv preprint arXiv:2311.13410},
  year = {2023}
}

@article{balgi2024deeplearning,
  author = {Balgi, Sourabh and Daoud, Adel and Pe{\~n}a, Jos{\'e} M. and Wodtke, Geoffrey and Zhou, Jesse},
  title = {Deep Learning With {DAGs}},
  journal = {Sociological Methods \& Research},
  year = {2025},
  doi = {10.1177/00491241251319291}
}

@inproceedings{balgi2022rho,
  author = {Balgi, Sourabh and Pe{\~n}a, Jos{\'e} M. and Daoud, Adel},
  title = {$\rho$-{GNF}: A Copula-based Sensitivity Analysis to Unobserved Confounding Using Normalizing Flows},
  booktitle = {Proceedings of the 11th International Conference on Probabilistic Graphical Models (PGM)},
  year = {2022},
  pages = {1--12}
}

\end{document}